\documentclass[a4paper,11pt]{article}
\usepackage{pos}

\title{Long-distance contributions to kaon decays on the lattice}
\author*[a]{En-Hung Chao}

\affiliation[a]{Physics Department, Columbia University,\\
New York City, New York 10027, USA}

\emailAdd{en-hung.chao@m4x.org}

\abstract{In the past decades, significant improvements have been made on standard-model predictions on kaon decays using lattice quantum chromodynamics.
In these proceedings, I review selected works on long-distance contributions to kaon decays and developments on QED corrections to those.
}

\FullConference{The 17th International Conference on Heavy Quarks and Leptons (HQL 2025)\\
15-19 September, 2025\\
Peking University, Beijing, China\\}

\newcommand{\kl}{K_{\rm L}}
\newcommand{\ks}{K_{\rm S}}

\begin{document}
\maketitle

\section{Kaon physics on the lattice}
Kaon decays provide great opportunities to probe standard-model (SM) physics and beyond.
Like many other flavor-changing processes, information on the elements of the Cabibbo-Kobayashi-Maskawa (CKM) matrix can be extracted from precisely measured decay modes of the kaons, allowing for detailed scrutiny of the charge-parity (CP) violation mechanisms described by the SM.
Rare kaon decays are of particular interest because of their sensitivities to New Physics, as flavor-changing neutral-current (FCNC) interactions are absent at tree-level in the SM.
Rare decay modes of the kaons which are mainly dominated by short-distance physics are identified as ``golden'' as they are theoretically very clean~\cite{Anzivino:2023bhp}. 
Precisely determining those decay modes from both theory and experiment can offer unique tests of the SM in the high-energy regime from low-energy measurements.
On the theory side, controlling non-perturbative hadronic effects is crucial for reaching the relevant precision targets.

Lattice quantum chromodynamics is a systematically-improvable method for estimating hadronic quantities from first principles. 
With algorithmic and technological advances, lattice QCD has become a competitive tool in the extraction of SM parameters from experiment~\cite{FlavourLatticeAveragingGroupFLAG:2021npn}.
In these proceedings, I will discuss a personal selection of recent efforts from the community into quantifying long-distance effects in kaon decays and the inclusion of QED corrections from lattice QCD.

\subsection{General strategies}
In an application of lattice QCD to kaon decays, one begins with an effective weak Hamiltonian where the heavy degrees of freedom above the scale of the mass of the charm quark are integrated out in perturbative theory.
Such an effective Hamiltonian is expressed as a linear combination of four-fermion operators multiplied by the associated Wilson coefficients.
Starting at a high energy scale where SM perturbation theory can be relied on, one performs renormalization by matching effective theories at different energy scales.
After running down to the scale where QCD becomes non-perturbative, lattice QCD comes into play and one needs to renormalize the same four-fermion operators defined on the lattice such that the same effective Hamiltonian is recovered in the continuum limit. 
As at high scales, the calculation of the Wilson coefficients involves continuum-theory regulators, one needs to work with a renormalization scheme at an intermediate scale which is defined for both lattice and continuum calculations to match the effective theories. 
A popular way to renormalize the lattice operators fulfilling this requirement is the \textit{regularization-invariant} (RI) scheme~\cite{Martinelli:1994ty}, which demands a set of spinor-projected amputated Green's function to match their tree-level values at a chosen kinematics.
An example of such is the symmetric momentum scheme (SMOM)~\cite{Sturm:2009kb}, where the two incoming quark momenta $p_1$ and $p_2$ satisfy the non-exceptional condition $p_1^2=p_2^2 = (p_1-p_2)^2= \mu^2$, with $\mu$ the renormalization scale.
In such a renormalization scheme, one needs to account for operator mixings due to the flavor-symmetry group and -- potentially -- the breaking of discrete symmetries due to lattice discretization.

As the name indicates, this renormalization scheme is independent of how the theory is regulated and can thus be used to obtain the Wilson coefficients for lattice operators by matching the renormalized lattice weak Hamiltonian to the continuum result at an energy scale $\mu$ where perturbative theory can still be trusted.

\section{Selected works on long-distance contributions and QED corrections}
In this section, I review a personal selection of works using lattice QCD to determine the contributions from long-distance effects at a length scale larger than the inverse of the charm quark mass and QED corrections. 

\subsection{$K^+\to\pi^+\nu\bar{\nu}$}
The processes of $K^i\to\pi^i\nu\bar{\nu}$ with $(K^i,\pi^i)=(K^+,\pi^+)$, $(\ks,\pi^0)$ and $(\kl, \pi^0)$ are dominated by short-distance physics as a result of the Glashow-Iliopoulos-Maiani (GIM) mechanism.
One expects perturbative calculations to give reliable predictions on those channels~\cite{Buras:2022wpw, DAmbrosio:2022kvb}.
Among those, the CP-violating $\kl\to\pi^0\nu\bar{\nu}$ process is considered as a golden mode as it mainly receives contribution from the top quark through direct CP-violation reflected in the CKM matrix. 
This decay might serve as a proxy to high energy physics from low energy experiments.
However, this process is not yet very well established experimentally ($\textrm{Br}(\kl\to\pi^0\nu\bar{\nu})\leq 2.2 \times 10^{-9}$ at the 90\% confidence level~\cite{KOTO:2024zbl}).
On the contrary, the CP-conserving $K^+\to\pi^+\nu\bar{\nu}$ has a clear experimental signal ($\textrm{Br}(K^+\to\pi^+\nu\bar{\nu})=13.0^{+3.3}_{-3.0}\times 10^{-11}$~\cite{NA62:2024pjp}) but receives a much larger contribution from the charm quark.
The long-distance contribution, which requires non-perturbative treatment of QCD, is estimated to be about 6\% of the total branching ratio estimated from perturbation theory~\cite{Isidori:2005xm}. 
Having a non-perturbative theory estimate to support this estimate is hence very well motivated.

A framework enabling lattice QCD calculations of the long-distance part of the $K^+\to\pi^+\nu\bar{\nu}$ decay amplitude has been worked out in Ref.~\cite{Christ:2016eae}.
In this framework, bi-local strangeness-changing operators are involved.
With a finite time extent $T$ as infrared regulator, upon Wick-rotating the relevant Minkowski-space correlation functions to Euclidean space, one will pick up terms containing a factor of $\exp(M_K-E_n)T$, with $E_n$ the energy of the intermediate state $|n\rangle$.
Due to the strangeness-changing operators, intermediate states less energetic than the kaon can be produced.
Those states propagate between the two local operators and introduce divergence as the infrared regulator is removed $T\to\infty$.
Fortunately, at fixed $T$, most of the matrix elements for the contributions from intermediate states which are less energetic than the kaon can be extract from standard lattice calculations. 
After reconstructing and subtracting those contributions, the resulting correlation function will have well-defined IR limit and can be extrapolated to the continuum and infinite-volume limit.

In Ref.~\cite{Christ:2019dxu}, the RBC/UKQCD group applied the aforementioned framework to a $32^3\times 64$, $N_{\rm f}=2+1$ flavor domain wall fermion ensemble at pion mass of $172$ MeV and kaon mass of $493$ MeV.
The valence charm quark in that work corresponds to an $\overline{\rm MS}$ mass of $m_c^{\overline{\rm MS}}(3\textrm{GeV})=750$ MeV.
Although at unphysical quark masses, the work suggests a mild momentum dependence of the decay amplitude and hints the smallness of the neglected unphysical $2\pi$ intermediate state contribution and the corresponding finite-volume effects, which paves the way for a calculation directly at physical quark masses.

\subsection{$K^+\to\pi^+ \mu^+ \mu^-$}
Unlike the $K^i\to\pi^i\nu\bar{\nu}$ cases, $K^i\to\pi^i \ell^+\ell^-$ decays with leptons $\ell=e,\mu$ are more sensitive to long-distance physics. 
Despite clear experimental signals~\cite{ParticleDataGroup:2024cfk}, the decay processes involving $K^+$ and $K_{\rm S}$ are dominated by long-distance physics as the GIM suppression is only logarithmic~\cite{Isidori:2005tv}. 
On the other hand, $K_{\rm L}\to\pi^0\ell^+\ell^-$, dominated by CP-violating processes, can provide cleaner theoretical constraints on different FCNC scenarios~\cite{Mescia:2006jd}.
Among those CP-violating mechanisms, the direct CP-violating part of $\kl\to\pi^0 \ell^+\ell^-$ is short-distance dominated.
In contrast, the indirect CP-violation due to $K^0-\bar{K}^0$ mixing requires a good determination of the one-photon exchange contribution to $K_{\rm S}\to\pi^0\ell^+\ell^-$.

Based on earlier work~\cite{Isidori:2005tv}, a framework using lattice QCD to compute the one-photon exchange contribution to $K^+\to\pi^+\ell^+\ell^-$ and $K_{\rm S}\to\pi^0\ell^+\ell^-$ is proposed in Ref.~\cite{Christ:2015aha}.
Analogous to the lattice framework to calculate $K^i\to\pi^i\nu\bar{\nu}$ discussed previously, one needs to remove the unphysical contributions from intermediate states which are less energetic than the kaon at each lattice ensemble before performing a continuum and infinite-volume extrapolation to the physical point.
In Ref.~\cite{RBC:2022ddw}, the RBC/UKQCD collaboration reports their results on $K^+\to\pi^+\ell^+\ell^-$ computed on a $48^3\times 96$ physical-pion-mass domain wall fermion ensemble at an inverse lattice spacing of $1.73$ GeV.
In this calculation, a conserved vector current is used in order to ensure U(1)$_V$ Ward identity. 
As a result, with a valence charm quark included, the overall result will be finite thanks to the GIM mechanism.
In their work, the data were generated for various light-than-physical charm quark masses at a fixed photon virtuality and extrapolated to the physical charm-quark mass. 
This calculation is however limited by the noisy quark-disconnected diagrams.
Promising work-in-progress on the implementation of new variance-reduction techniques to overcome this difficulty was presented at the Lattice 2024 Conference~\cite{Hodgson:2025iit}.

\subsection{$K_{\rm L}\to \mu^+\mu^-$}
Compared to the aforementioned decay channels, the rare $K_{\rm L}\to \mu^+\mu^-$ decay is very precisely measured, with a branching ratio of $6.84(11)\times 10^{-9}$~\cite{E871:2000wvm}.
It was suggested to exploit this process as a benchmark for FCNC processes.
Nevertheless, while the O$(G_{\rm F}^2)$ short-distance contribution is well determined from perturbation theory~\cite{Gorbahn:2006bm}, the O($G_{\rm F}\alpha_{\rm QED}^2$) two-photon exchange contribution is expected to be non-negligible for the targeted precision. 
For the latter, non-perturbative methods are needed in order to treat the hadronic effects rigorously. 
In particular, a first-principles calculation is strongly motivated as the interference between the O($G_{\rm F}^2$) and O($G_{\rm F}\alpha_{\rm QED}^2$) contributions plays a crucial role in the determination of the total decay amplitude. 

A lattice QCD framework for calculating the two-photon exchange contribution to $\kl\to\mu^+\mu^-$ was first proposed in Ref.~\cite{Christ:2020bzb}.
This method is inspired by the recent success of lattice calculations of the hadronic light-by-light scattering contribution to the anomalous momentum of the muon, where the QED${}_{\infty}$ formalism with the photon propagators treated in the continuum and infinite volume was used~\cite{Asmussen:2016lse, Blum:2017cer}.
Similar to the lattice rare kaon applications reported earlier, unphysical, divergent contributions also arise upon Wick-rotating the Minkowski-space amplitude.
The present situation is a bit more delicate due to the presence of the two-pion intermediate states whose energies are below the mass of the kaon at rest, whose contributions are more difficult to reconstruct with well-controlled finite-volume effects.
Based on phenomenology-inspired models, the corresponding systematic error is estimated to be at the ten-percent level at worst~\cite{Chao:2024vvl}.
In spite of not being able to precisely resolve the low-energy two-pion intermediate state contributions within the proposed framework, being able to attain the projected precision will already make significant impact in ameliorating the SM prediction.

The first numerical results from an exploratory study on a $24^3\times 64$ physical-pion mass ensemble at an inverse lattice spacing of $1.023$ GeV from the RBC/UKQCD collaboration are reported in Ref.~\cite{Boyle:2025fug}.
The obtained results can however not yet be compared to experimental results as extra counter terms are required for renormalization under their 2+1 quark flavor setup, where the GIM cancellation is absent. 

\subsection{Radiative corrections to $\pi_{\ell 2}$ and $K_{\ell 2}$}
When sub-percent level precisions are targeted, one needs to include both the O$(\alpha_{\rm QED})$ leading electromagnetic and O$((m_d-m_u)/\Lambda_{\rm QCD})$ strong isospin-breaking corrections. 
A lattice formulation of QED is incompatible with periodic boundary conditions as the Gau\ss{} law does not allow non-zero static electric charge in such a box.
Different formalisms have been developed, such as QED${}_{\rm L}$~\cite{Hayakawa:2008an}, QED${}_{\rm m}$~\cite{Endres:2015gda}, QED${}_{\rm C}$~\cite{Lucini:2015hfa} and QED${}_{\infty}$ mentioned earlier.

The ratio of the CKM matrix elements $|V_{ud}|/|V_{us}|$ can be extracted from accurate experimental measurements of the ratio of the $K_{\ell 2}$ and $\pi_{\ell 2}$ decay rates, $\Gamma(K^\pm\to\ell^\pm \nu_\ell)/\Gamma(\pi^\pm\to\ell^\pm \nu_\ell)$, provided that the isospin-breaking sensitive ratio $f_{K^\pm}/f_{\pi^\pm}$ is precisely determined from theory.
When it comes to QED corrections, one must account for the radiative corrections contributing at the same order in $\alpha_{\rm QED}$ in order for the combined decay rate to be IR-convergent (Bloch-Nordsieck theorem).
A lattice-QCD framework accounting for such divergence-cancellations in semi-leptonic decays was proposed by the RM123-SOTON collaboration~\cite{Carrasco:2015xwa}, where the QED corrections and strong-isospin breaking effects are included by a perturbative expansion in $\alpha_{\rm QED}$ and $(m_d-m_u)/\Lambda_{\rm QCD}$~\cite{deDivitiis:2013xla}.
In this framework, the $P_{\ell 2}$ decay rate with $P=K,\pi$ is decomposed into two well-defined quantity
\begin{equation}
\Gamma(P_{\ell 2}) = \lim_{L\to \infty}\left[
\Gamma_0(L) - \Gamma^{\rm pt}(L)
\right] + \lim_{\mu_\gamma\to 0}\left[
\Gamma^{\rm pt}_0(\mu_\gamma) + \Gamma_1^{\rm pt}(\Delta E_\gamma, \mu_\gamma)
\right]\,,
\end{equation}  
where $\Gamma_0$ is the decay rate with virtual photon corrections, $\Gamma^{\rm pt}$ is the universal IR-divergent which can be calculated with a point-like interaction and $\Gamma_{1}^{\rm pt}$ is the rate with real photon corrections estimated with a point-like interaction up to photon energy $\Delta E_{ \gamma}$. 
Both pieces are IR-convergent and can thus be computed separately and combined with the IR regulator removed. 
Based on this framework, two lattice QCD estimates for the relative correction $\delta R_{K\pi}$ to the ratio are now available from the RM123-SOTON collaboration~\cite{DiCarlo:2019thl} ($R_{K\pi}^{\rm RM123S}=-0.0126(14)$) and the RBC/UKQCD collaboration~\cite{Boyle:2022lsi} ($R_{K\pi}^{\rm RBC/UKQCD} = -0.0086(13)(39)_{\rm vol}$ where the systematic error due to finite-volume effects are isolated from the rest combined).
These estimates are compatible with the prediction from Chiral Perturbation theory~\cite{Cirigliano:2011tm}.

Further developments and improvements on the radiative corrections to pseudoscalar meson decays are being pursued vividly. 
From the RM123-SOTON collaboration, form factors for $K\to\ell\nu_{\ell}\gamma$ with $\ell = e,\mu$ are calculated directly from lattice QCD~\cite{DiPalma:2025iud}, which are compared to the experimental results and provide insights to the existing tensions between experiments.
A framework to calculate radiative leptonic decays of pseudoscalar meson $P\to \ell \nu_{\ell }\gamma$ with exponentially-suppressed finite-volume effects based on the infinite-volume reconstruction (IVR) method~\cite{Feng:2018qpx} has been proposed by the RBC/UKQCD collaboration~\cite{Christ:2023lcc}.\footnote{Shortly after the HQL2025 Conference, first numerical results based on the proposed method became available~\cite{Christ:2025ufc}.}
In particular, the IVR method is applicable to Feynman diagrams where a photon is attached to a hadronic external state.
Prospects of calculating $\bar{K}^0\to\pi^+\ell^-\nu_{\ell}$ based on the same formalism have also been discussed Refs.~\cite{Christ:2023lcc,Christ:2024xzj}.

\acknowledgments
E.-H.C. was supported in part by the U.S. Department of Energy (DOE) grant No. DE-SC0011941.

\end{document}